\documentclass[twocolumn,aps,prl]{revtex4}
\usepackage{mathrsfs}
\usepackage{amsmath}
\usepackage{amsfonts}
\usepackage{amssymb}
\usepackage{amsthm}
\usepackage{graphicx}
\begin{document}

\title{A theory for magnetic-field effects of nonmagnetic
organic semiconducting materials}
\author{X. R. Wang}
\affiliation{Physics Department, The Hong Kong University of Science
and Technology, Clear Water Bay, Hong Kong SAR, China}
\author{S. J. Xie}
\affiliation{School of Physics, National Key Laboratory of Crystal
Materials, Shandong University, Jinan, P. R. China}

\begin{abstract}
A universal mechanism for strong magnetic-field effects of
nonmagnetic organic semiconductors is presented. A weak magnetic
field (less than hundreds mT) can substantially change the charge
carrier hopping coefficient between two neighboring organic
molecules when the two hopping states are not too symmetric. 
Under the illumination of lights or under a high electric field, 
the change of hopping coefficients leads also to the change of 
polaron density so that photocurrent, photoluminescence,
electroluminescence, magnetoresistance and electrical-injection
current become sensitive to a weak magnetic field.
The present theory can not only explain all observed features,
but also provide a solid theoretical basis for the widely used
empirical fitting formulas.
\end{abstract}
\date{\today}
\maketitle
One of the long-term\cite{merrifield,williams,wohlgenannt,hu} 
unsolved fundamental issues in organic physics is the mechanism 
behind the strong responses of electrical and optical properties 
of nonmagnetic organic semiconductors to a weak magnetic field, 
known as organic magnetic-field effect (OMFE). The recent revival 
interest in OMFE 
of organic semiconductors is largely due to its importance 
in fundamental science and technology applications\cite{hu}. 
Firstly, there is a belief that the OMFE can be used as a 
powerful tool to probe microscopic processes of organic materials. 
Secondly, the OMFE can be used to develop new multifunctional 
organic devices\cite{shi}. Experiments showed that OMFE has 
following surprising yet universal features. 1) The OMFE appears 
in vast different organic semiconductors without any magnetic 
elements at room temperature although the possible energy level 
shifts due to the presence of a magnetic field are orders 
magnitude smaller than the thermal energy and other energy scales. 
2) The electroluminescence, photocurrent, photoluminescence, and
electrical-injection current are very sensitive to weak magnetic
field with both positive and negative OMFE though positive OMFE 
(or negative magnetoresistance (MR) in the convention terminology) 
at very weak field is typically observed. 3) The OMFE can 
often be fitted by two empirical formulas: $[B/(B+B_0)]^2$ and
$B^2/(B^2+B_0^2)$\cite{Robbert}, where $B$ is the applied magnetic 
field. In the theoretical side, it is known\cite{wohlgenannt,hu} 
that familiar MR mechanisms such as Lorentz force, conventional 
hopping MR, electron-electron interaction and weak localization 
are highly unlikely to be the cause behind the OMFE. 
The current belief in the community is that the OMFE is 
intimately tied to spin physics involving spin configuration, 
spin correlation, and spin flip\cite{hu}. However, there is no 
convincing arguments why an extremely small Zeeman energy can 
beat other much larger energy scales in controlling electron 
spin dynamics to generate this OMFE. Both extensive experimental 
and theoretical studies so far are suggesting that a novel 
explanation is needed. This new MR mechanism should explain 
not only all OMFE features, but also why the similar effects 
do not often appear in the usual inorganic semiconductors. 
In this report, we present such a theory that does not 
explicitly rely on the electron spin degrees of freedom. 
It is showed that the OMFE originates from the substantial 
change of electron hopping coefficient in a magnetic field 
because of narrow bandwidth of organic semiconductors and 
asymmetry in organic molecules. 

Organic semiconductors have a few distinct properties that their 
inorganic counterparts do not have. Firstly, unlike an atom that 
is sphere-like, an organic molecule is highly irregular. As a 
result, organic molecular wavefunction has no obvious symmetry. 
Secondly, organic molecules in organic semiconductors are bonded 
by the Van der Waals force so that their bands are very narrow in 
comparison with an order of 10eV bandwidth for their inorganic 
counterparts \cite{Troisi}. Thirdly, the intramolecular excitons 
have strong binding energies of order of eV\cite{hu}. 
On the other hand, the electron and hole become polaron pair 
when they are located on different molecules because of 
weak intermolecular exciton binding energy\cite{hu}.  
The electrical properties of an organic semiconductor are 
mainly determined by the motion of polarons since the motion 
of excitons does not contribute to the electric current.
The singlet excitons are responsible to the luminescence.  
A weak field should not change much of energy levels of various 
excitation states so that their populations at thermal 
equilibrium are not sensitive to a magnetic field because they 
depend only on the energy level distribution and the temperature. 
Any significant change in magnetoresistance near the 
quasiequilibrium state must come from the mobility change.
The question is whether a weak field of 100mT can change 
the mobility of polarons in organic semiconductors. 
In the usual inorganic crystals with s-like wavefunction, 
the answer is no. However, we argue below that this can 
indeed happen in organic conjugated materials with highly 
irregular molecular wavefunction. 

\begin{figure}[tbph]
\begin{center}
\includegraphics[width=0.9\columnwidth]{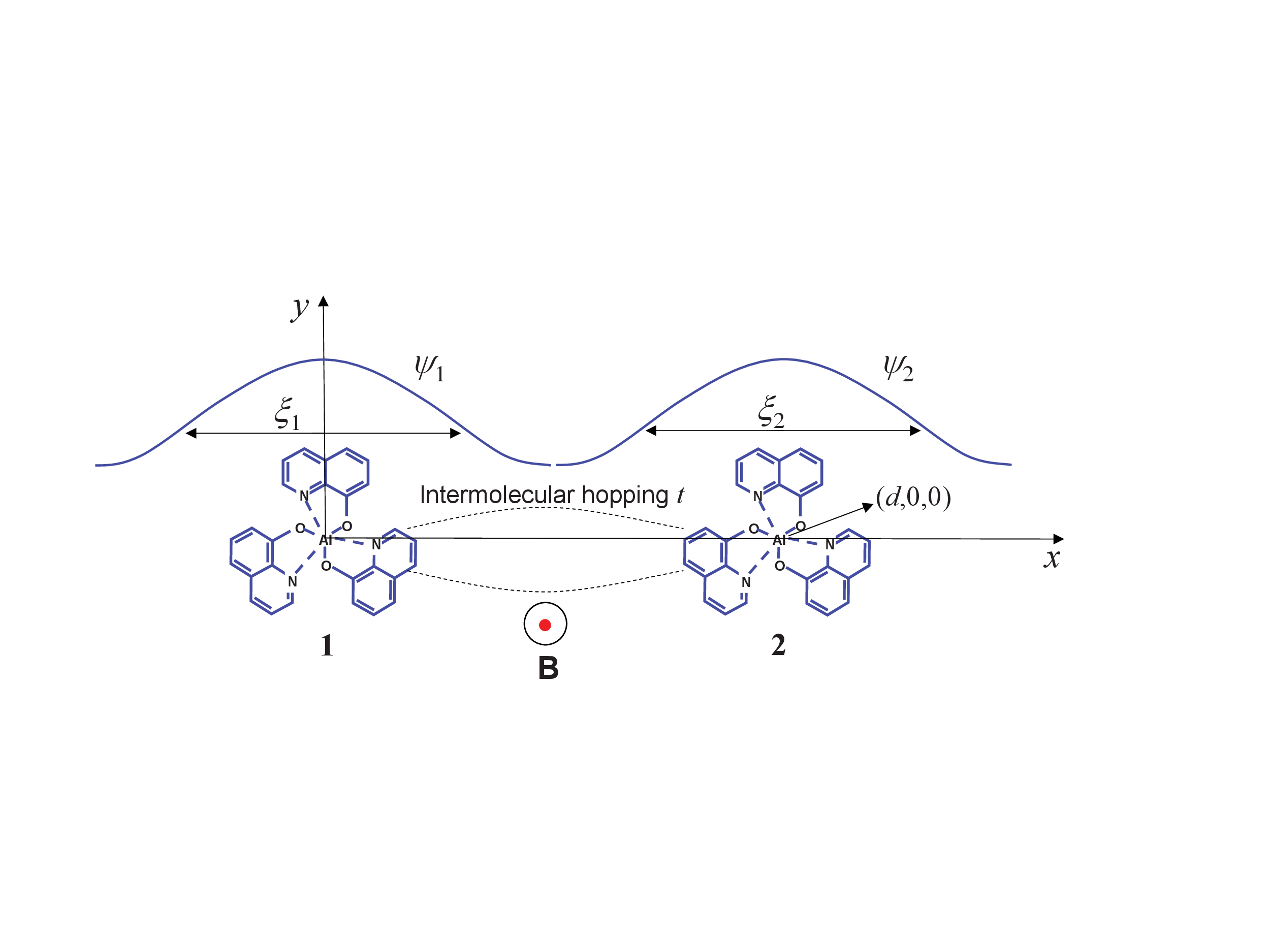}
\end{center}
\caption{Schematical draw of two organic molecules (Alq$_3$) 
separated by a distance $d$ and aligned along x-direction. 
Molecule {\bf 1} is centered at the origin, and molecule 
{\bf 2} is centered at $(d,0,0)$. The field is assumed to be 
along z-direction. $\psi_1$ and $\psi_2$ are two localized 
states with localization lengths $\xi_1$ and $\xi_2$ on 
molecules {\bf 1} and {\bf 2}, respectively. }
\label{fig1}
\end{figure}

In order to understand why a weak magnetic field can change charge
carrier (electron and hole or polaron) mobility in an organic
conjugated material, we consider a system with two molecules
separated by a distance $d$ as schematically shown in Fig. 1.
One-electron Hamiltonian in a magnetic field can in general be
described by
\begin{equation}
\label{Hamiltonian} H=
-\frac{1}{2m}(\vec{p}-\frac{e}{c}\vec{A})^{2}+ V_1+V_2
\end{equation}
where $V_1$ and $V_2$ are the potential created by molecules {\bf 1}
and {\bf 2}, respectively. $\vec A$ is the vector potential due to
magnetic field $\vec B$. For the simplicity and clarity, we shall
assume that the two molecules are aligned along x-direction, the
field is along the z-direction (pointing out of the paper). 
The important quantity for electron transport is the tunneling 
matrix element between two molecules. When an electron tunnels 
from an initially occupied state, say $\psi_1$ of molecule 
$\bf 1$, to empty state $\psi_2$ of molecule $\bf 2$ with 
tunneling matrix $t$ , it will contribute to the hopping 
probability $P$ (per unit time), proportional to $|t|^2\exp
(-\Delta\epsilon_{12}/(KT))$, where $\Delta \epsilon_{12}$
describes the relative energy level with respect to the Fermi
level\cite{boris}. The hopping conduction can be regarded
as an electron diffusion process with a diffusion constant 
$D=Pd^{2}$, where $d$ is the average distance between two 
neighboring molecules. According to the Einstein relation, 
the electron mobility $\mu$ is given by $\mu=eD/(KT)$ which is 
related to the conductivity in the conventional way\cite{boris}. 
Therefore, we can concentrate on how the tunneling matrix 
element depends on the magnetic field in order to study the 
magnetoresistance of the system.

In the tight-binding approximation\cite{landau}, one of the authors
in an early publication\cite{xrw} has generalized the Bardeen's
transfer matrix formalism to high dimension and in the presence of a
magnetic field. In 3D, it is
\begin{equation}
\label{hopping} t=\frac{\hbar^2}{m}\int [ (\psi^\star_1 \frac
{\partial \psi_{2}}{\partial x} - \psi_{2}\frac
{\partial\psi^\star_1} {\partial x}) -
\frac{2i}{\phi_0}(\vec{A}\cdot \hat{x})
\psi^{\star}_{1}\psi_{2}]|_{x=\frac{d}{2}}dydz,
\end{equation}
where $\phi_0=c\hbar /e$ is the flux quanta. 
For small $\vec{A}$ when the magnetic length
$l_B=\sqrt{\phi_0/B}$ is bigger than $d$, magnetic confinement
that is responsible for the exponential increase of resistance in
the usual hopping conduction can be neglected and $\psi_1$ and
$\psi_2$ do not depend on $B$ to the zero order approximation. 
Then the magnitude of the field-independent part of $t$ is 
order of $$\frac{\hbar^2} {m\xi}\int\psi^\star_1\psi_2|_{x=
\frac{d}{2}}dydz$$ while that of the field dependent part 
is $$\frac{\hbar^2}{m}\frac {1}{l_B^2}\int y\psi^\star_1
\psi_2|_{x=\frac{d}{2}} dydz.$$ $\xi^{-1}=\xi_2^{-1}+
\xi_1^{-1}$, and $\xi_1$ and $\xi_2$ are the localization 
lengths of $\psi_1$ and $\psi_2$, respectively. 
Both terms depend on the nature of the wavefunctions. 
This explains why the OMFE value varys from sample to sample, 
and from material to material\cite{hu}. The second term 
vanishes for s-like wavefunction. This is why the similar 
phenomena do not show up in usual inorganic semiconductors. 
Due to the irregular structures of organic molecules, 
one will expect an appreciable value for the second term, 
resulting in a sizable change of the hopping probability. 

Due to the Van der Waals bonding, the OMFE is measurable 
only under an optical injection of carriers or an electric 
carrier injection by an electric field above a threshold. 
When an organic semiconductor is under the illumination of 
a light or under a high electric field, the field dependent 
$t$ results in a field dependence of polaron density. 
Take optical injection of carriers as an example, under the 
illumination of a light, an electron in a highest occupied
molecular orbit (HOMO) absorbs a photon and jumps to a higher empty
molecular orbit of the same molecule. As schematically illustrated
in Fig. 2, the excited electron can either dump its excessive
kinetic energy to its environment and forms an exciton with the hole 
left behind or jumps to neighboring molecules and becomes polarons. 
Depending on the relative probabilities of excited electrons (holes) 
staying in the same molecules and jumping to different molecules, 
the polaron density shall vary with the illumination intensity. 
Let us denote the probability (per unit time) of a pair of electron 
and hole on the same molecule forming an exciton by $P_0\sim \hbar/\tau$,
where $\tau$ is the typical time for a pair of electron and hole to
form an exciton. $P_0$ is not sensitive to a weak field since the
field cannot change much molecule orbits that determine $P_0$.
Then the polaron generation rate per unit volume is $JP/(P_0+P)$
where $J$ is the photon absorption rate per unit volume
and $P\propto |t|^2$ is the intermolecular hopping probability.
Without the illumination of a light, polaron density shall reach 
its equilibrium density $n_0$ at a rate of $\gamma (n-n_0)$, 
where $\gamma$ is polaron decay rate. At balance, $JP/(P_0+P)=
\gamma (n-n_0)$, thus the photon-generated polaron density 
$n$ should be $n_0+JP/[\gamma(P_0+P)]$. Clearly, $B$-dependence 
of $P$ results in a $B-$dependence of polaron density.
\begin{figure}[tbph]
\begin{center}
\includegraphics[width=0.9\columnwidth]{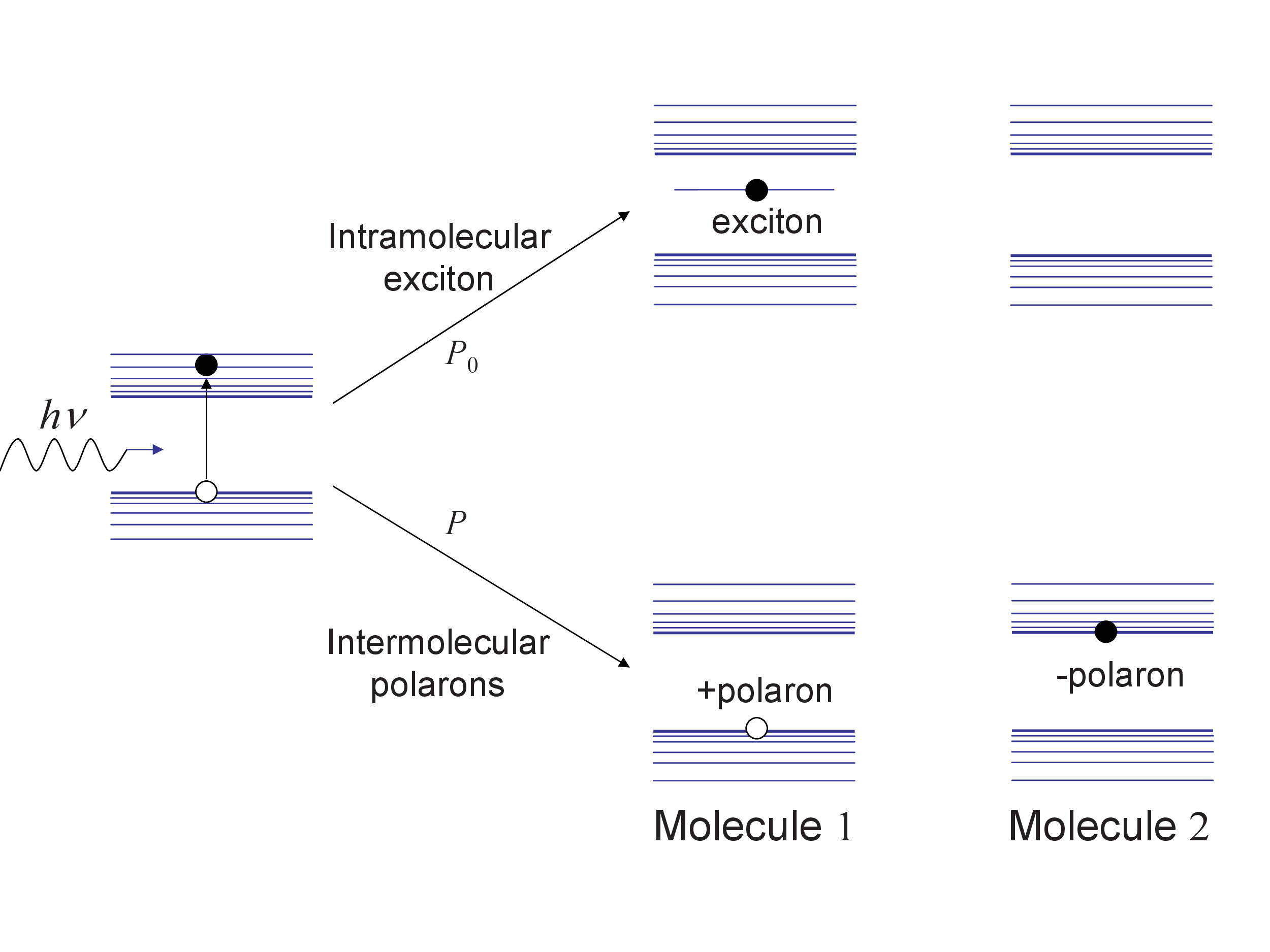}
\end{center}
\caption{Schematic illustration of polaron and exciton formation
after a pair of electron and hole is created by a photon absorption.
The excited electron-hole pair has probability $P$ jumping to the 
neighboring molecules to form positive-charged and negative-charged 
polarons, and probability $P_0$ to form an exciton. 
} \label{fig2}
\end{figure}

Since the OMFE can often be fitted by two empirical functions 
$B^2/(B^2+B_0^2)$ and $[B/(B+B_0)]^2$\cite{Robbert}, a correct 
theory should be able to provide a theortical basis for the fact. 
According to Eq. \eqref{hopping}, $t$ takes a form of $B_0+iaB$ 
with $B_0$ and $a$ real and field-independent parameters if 
$\psi_1$ and $\psi_2$ are real functions. This is the case when 
the molecule orbits involved in hopping are localized or not 
degenerated\cite{landau}. In this case, $P\propto|t|^2=a(B^2+
B_0^2)$ and the polaron density shall depend on the magnetic 
field as 
$\frac{P}{\gamma(P_0+P)}J+n_0=n_0'+\alpha B^2/(B^2+B_0^2)$, 
where $n_0$, $n_0'$, $\alpha$ and $B_0$ are B-independent
parameters that depend on the molecule orbits involved.
Thus, $B^2/(B^2+B_0^2)$ is a natural OMFE function for 
$t=B_0+iaB$. 
The second type of empirical function appears naturally for 
$t=i(B_0+aB)$. According to Eq. \eqref{hopping}, this can 
happen when the spatial derivatives of $\psi_1$ or $\psi_2$ 
are the functions multiplied by pure imaginary numbers. 
Of course, this must correspond to degenerated states.
In this case, the leading term in the polaron density takes a
form of $[B/(B+B_0)]^2$ in a similar argument when $P\gg P_0$.
In reality, electron (polaron) hopping between two organic
molecules should involve many molecule orbits, especially in
photophysical processes and in a high electric field.
One then needs to add contributions from all hopping events.
Thus, it is likely that both $B^2/(B^2+B_0^2)$ and
$[B/(B+B_0)]^2$ processes are presented, and OMFE should
then be fitted by the linear combinations of these
two functions, consistent with experimental findings.

The novel mechanism is very robust. At the room temperature, 
the transport of charge carriers will involve many different 
molecule orbits. Each hopping event will subject to the 
influence of this mechanism as long as magnetic confinement is 
negligible ($l_B>d$) and molecular structure is not sphere-like. 
Of course, thermal average over all hopping events is needed. 
Molecule-molecule orientation in organic semiconductors should 
be quasi-random due to the nature of organic molecules. 
A magnetic field can be along any direction with respect to 
the molecule-molecule bond instead of perpendicular direction  
as assumed in the above discussion. This explains why OMFE is 
not sensitive to the field direction in devices. According to 
Eq. \eqref{hopping}, different angle between the field and 
molecule-molecule bond leads to different hopping coefficient. 
It should also be emphasized that the mechanism present here 
does not depend on electron spins, and it does not require 
large energy splits of different spin configurations.
It is applicable to both bipolar and hole-only (or 
electron-only) devices. Differ from the previous theories 
that try to relate the OMFE to the changes of electron levels, 
the present theory attributes the OMFE to the change of 
electron hopping coefficient in a field. 
Thus, it does not have all the troubles as those spin-dynamics
related theories involve concepts of excitons and bipolarons
\cite{Kalinowski,Prigodin,Hu2,Robbert,Majumdar}. 

In conclusion, we present a novel mechanism for the OMFE for
nonmagnetic organic semiconductors. The mechanism is very general
and robust for organic semiconductors, but is normally not 
important for usual covalently bonded inorganic semiconductors. 
The mechanism can not only explain all experimentally observed 
OMFE, but also provides a solid theoretical basis for the  
empirical OMFE formulas. New experiments are needed to firmly 
establish this mechanism as the genuine cause of the OMFE.

This work is supported by Hong Kong UGC grants (\#604109,
HKU10/CRF/08-HKUST17/CRF/08, and RPC07/08.SC03). 
SJX is supported by the National Basic Research Program 
of China (Grant No.2009CB929204 and No.2010CB923402) and the 
NNSF of China (Grant No. 10874100).

\end{document}